\documentclass{jpp}
\usepackage[T1]{fontenc}
\usepackage[latin9]{inputenc}
\usepackage{graphicx}
\usepackage{natbib}
\usepackage{amsmath}
\usepackage{amssymb}

\ifCUPmtlplainloaded \else
  \checkfont{eurm10}
  \iffontfound
    \IfFileExists{upmath.sty}
      {\typeout{^^JFound AMS Euler Roman fonts on the system,
                   using the 'upmath' package.^^J}%
       \usepackage{upmath}}
      {\typeout{^^JFound AMS Euler Roman fonts on the system, but you
                   dont seem to have the}%
       \typeout{'upmath' package installed. JPP.cls can take advantage
                 of these fonts, if you use 'upmath' package.^^J}%
      }
  \else
  \fi
\fi

\ifCUPmtlplainloaded \else
  \checkfont{msam10}
  \iffontfound
    \IfFileExists{amssymb.sty}
      {\typeout{^^JFound AMS Symbol fonts on the system, using the
                'amssymb' package.^^J}%
       \usepackage{amssymb}%
         
         \let\geq=\geqslant
      }{}
  \fi
\fi

\ifCUPmtlplainloaded \else
  \IfFileExists{amsbsy.sty}
    {\typeout{^^JFound the 'amsbsy' package on the system, using it.^^J}%
     \usepackage{amsbsy}}
    {}
\fi

\newsavebox{\astrutbox}
\sbox{\astrutbox}{\rule[-5pt]{0pt}{20pt}}

\title{Linear Collisionless Landau Damping in Hilbert Space}

\author{Alessandro Zocco}

\affiliation{Max-Planck-Institut für Plasmaphysik, Wendelsteinstrasse, D-17489,
Greifswald, Germany}

\begin{document}
\maketitle
\begin{abstract}
The equivalence between the Laplace transform {[}Landau L., J. Phys.
USSR \textbf{10} (1946), 25{]} and Hermite transform {[}Zocco and
Schekochihin, Phys. Plasmas \textbf{18}, 102309 (2011){]} solutions
of the linear collisionless Landau damping problem is proven. 
\end{abstract}

\section{Hermite revival }

Vibrations in plasmas can be damped even in the absence of collisions.
This phenomenon is known as Landau damping \citep{landau}. Landau
damping acts on different types of waves: Langmuir waves, sound waves,
kinetic Alf\'en waves, drift waves, and many more. It basically occurs
any time the momentum associated to a wave propagating in a plasma
can sample regions of velocity-space where the plasma distribution
function is prone to the formation of a singularity. A particularly
interesting case is that of kinetic Alfv\'en waves (KAW) \citep{PhysRevLett.35.370},
and will be studied here. These waves are of pivotal importance in
many physical phenomena in magnetised plasmas such as magnetic reconnection
\citep{ReviewReconn}, auroral electromagnetic turbulence \citep{observationaurora},
and astrophysical gyrokinetics \citep{alex}, for instance.

A simple hybrid fluid-kinetic model that supports kinetic Alfv\'en
waves \textsl{and } their Landau damping was introduced by  \citet{zocco:102309}.
There, an efficient way to simulate the model equations numerically
via the Hermite representation of velocity-space was proposed. This
proved to be useful to study electron heating and nonlinear Landau
damping of kinetic Alfv\'en waves in collisionless magnetic reconnection
\citep{loureiro-scecco-zocco}.

While some physical insight on nonlinear Landau damping can be obtained
by ``brute force'' numerical simulations, \citep{loureiro-scecco-zocco},
it is still unclear what is the relation between the original linear
result of \citet{landau} and the Hermite representation of
velocity-space \citep{hammet-hermite,smith,sugama:2617,zocco:102309}.
In this work we address this issue, and prove the equivalence of the
two treatments.

\section{Equations}

We briefly review the system of equations studied. Details can be
found in \citet{zocco:102309}. We start with the collisionless
electron drift-kinetic equation \citep{frieman:502} for $h_{e}=F_{e}-F_{0e}(1+e\varphi/T_{0e})$

\begin{equation}
\begin{split} & \frac{\partial h_{e}}{\partial t}+\mathbf{v}_{E}\cdot\nabla h_{e}+v_{\parallel}\hat{\mathbf{b}}\cdot\nabla h_{e}=-\frac{eF_{0e}}{T_{0e}}\frac{\partial}{\partial t}\left(\varphi-\frac{v_{\parallel}A_{\parallel}}{c}\right)\end{split}
,\label{eq:edk}
\end{equation}
where $\mathbf{v}_{E}=cB_{0}^{-1}(-\partial_{y}\varphi\mathbf{e}_{x}+\partial_{x}\varphi\mathbf{e}_{y})$
is the $\mathbf{E}\times\mathbf{B}$ drift velocity, $\hat{\mathbf{b}}\cdot\nabla=\partial_{z}-B_{0}^{\text{-1}}\{A_{\parallel},\},$
$\varphi$ and $A_{\parallel}$ are the electrostatic and magnetic
potential, $\{,\}$ is the Poisson bracket, and 
\begin{equation}
F_{0e}=\frac{n_{0e}}{\left[\pi v_{the}^{2}\right]^{3/2}}e^{-\frac{v_{\parallel}^{2}+v_{\perp}^{2}}{v_{the}^{2}}}
\end{equation}
is the Maxwellian equilibrium with temperature $T_{0e}=m_{e}v_{the}^{2}/2.$
Equation (\ref{eq:edk}) describes the statistical properties of a
magnetised electron species for low-frequency anisotropic fluctuations
in the presence of a mean magnetic field. Here, this is a constant
straight magnetic field $\mathbf{B}_{0}=B_{0}\hat{\mathbf{b}}$, whose
direction defines the z axis. 

We introduce a formal mass ratio
expansion for the electron perturbed distribution function, so that
to zeroth order 
\begin{equation}
h_{e}=\left(-\frac{e\varphi}{T_{0e}}+\frac{\delta n_{e}}{n_{0e}}+\frac{v_{\parallel}u_{\parallel e}}{T_{0e}}m_{e}\right)F_{0e}+g_{e}+\mathcal{O}\left(\frac{m_{e}}{m_{i}}\right),\label{eq:split}
\end{equation}
here $\delta n_{e}=\int d^{3}\mathbf{v}h_{e}$, $u_{\parallel e}=n_{0e}^{-1}\int d^{3}\mathbf{v}v_{\parallel}h_{e}$,
and 
\begin{equation}
\int d^{3}\mathbf{v}(1,v_{\parallel})g_{e}\equiv0.\label{eq:vincoli}
\end{equation}
Using expression (\ref{eq:split}) in Eq. (\ref{eq:edk}), and taking
the zeroth moment we obtain the electron continuity equation 
\begin{equation}
\frac{d}{dt}\frac{\delta n_{e}}{n_{0e}}=-\hat{\mathbf{b}}\cdot\nabla u_{\parallel e}\label{eq:elcontapp}
\end{equation}
 with $d/dt=\partial_{t}+\mathbf{v}_{E}\cdot\nabla.$ 

The first moment of Eq. (\ref{eq:edk}) yields the generalized Ohm's
law 
\begin{equation}
\begin{split} & \frac{d}{dt}(A_{\parallel}-d_{e}^{2}\nabla_{\perp}^{2}A_{\parallel})=-c\frac{\partial\varphi}{\partial z}+\frac{T_{0e}c}{e}\hat{\mathbf{b}}\cdot\nabla\left[\frac{\delta n_{e}}{n_{0e}}+\frac{\delta T_{\parallel e}}{T_{0e}}\right]+\frac{m_{e}c}{e}\frac{1}{n_{oe}}\frac{d}{dt}n_{0e}u_{\parallel i},\end{split}
\label{eq:grnohmslawapp}
\end{equation}
 with 

\begin{equation}
\frac{\delta T_{\parallel e}}{T_{0e}}\equiv\frac{1}{n_{0e}}\int d^{3}\mathbf{v}2\frac{v_{\parallel}^{2}}{v_{the}^{2}}g_{e}.
\end{equation}
 In Eq. (\ref{eq:grnohmslawapp}), we used parallel Ampere's law 
\begin{equation}
u_{\parallel e}=\frac{e}{m_{e}c}d_{e}^{2}\nabla_{\perp}^{2}A_{\parallel}+u_{\parallel i}
\end{equation}
where $d_{e}=c/\omega_{pe}$ is the electron skin depth, and $\omega_{pe}$
the electron plasma frequency. An equation for $g_{e}$ is derived
after using Eqs. (\ref{eq:elcontapp}) and (\ref{eq:grnohmslawapp})
in Eq. (\ref{eq:edk}). The result is 
\begin{equation}
\begin{split} & \frac{dg_{e}}{dt}+v_{\parallel}\left[\hat{\mathbf{b}}\cdot\nabla g_{e}-F_{0e}\hat{\mathbf{b}}\cdot\nabla\frac{\delta T_{\parallel e}}{T_{0e}}\right]=F_{0e}\left(1-2\frac{v_{\parallel}^{2}}{v_{the}^{2}}\right)\hat{\mathbf{b}}\cdot\nabla\left[\left(\frac{e}{m_{e}c}d_{e}^{2}\nabla_{\perp}^{2}A_{\parallel}+u_{\parallel i}\right)\right].\end{split}
\label{eq:krehminhom}
\end{equation}

The kinetic information is embedded in the function 
\begin{equation}
\frac{\delta T_{\parallel e}}{T_{0e}}\equiv\frac{1}{n_{0e}}\int d^{3}\mathbf{v}2\frac{v_{\parallel}^{2}}{v_{the}^{2}}g_{e}.
\end{equation}
The system of equations is closed by solving for the ion dynamics,
imposing quasineutrality \citep{zocco:102309} 
\begin{equation}
\frac{\delta n_{e}}{n_{0e}}=\frac{\delta n_{i}}{n_{0i}},
\end{equation}
and using the ion solution\citep{PhysRevLett.42.1058,antonsen-coppi,0029-5515-22-1-005,pegoraro:364,PhysRevLett.66.425,zocco:102309,0741-3335-54-3-035003}
\begin{equation}
\frac{\delta n_{i}}{n_{0i}}=\int_{-\infty}^{+\infty}dpe^{ipx}F(p\rho_{i})\frac{Ze\varphi}{T_{0i}}\equiv\hat{F}\frac{Ze\varphi_{\mathbf{k}}}{T_{0i}},\label{eq:GKPL}
\end{equation}
 with 
\begin{equation}
F(p\rho_{i})=-(1-\Gamma_{0}),
\end{equation}
$\tau=T_{0i}/T_{0e},$ $Z$ is the charge number, $k^{2}=k_{x}^{2}+k_{y}^{2},$
and $\Gamma_{n}=\exp[-p^{2}\rho_{i}^{2}/2]I_{n}(p^{2}\rho_{i}^{2}/2),$
where $I_{n}$ is the modified Bessel function \citep{abram}. The
``hat'' on $F(p\rho_{i})$ is a short-hand notation for the inverse
Fourier transform.The ion solution garantees that $u_{\parallel i}=0.$
In the truly collisionless case cosidered here, we can solve the
electron kinetic equation (\ref{eq:krehminhom}). We linearize Eq.
(\ref{eq:krehminhom}) using perturbations of the form $g_{e}\propto\exp[-i\omega t+ik_{\parallel}v_{the}].$

After using standard algebra, we find \citep{zocco:102309},
\begin{equation}
\frac{\delta T_{\parallel e}}{T_{0e}}=-\frac{k_{\parallel}}{\left|k_{\parallel}\right|}\frac{u_{\parallel e}}{v_{the}}\frac{Z(\zeta)-2\zeta\left[1+\zeta Z(\zeta)\right]}{1+\zeta Z(\zeta)},\label{eq:Tkin}
\end{equation}
where $\zeta=\omega/(\left|k_{\text{\ensuremath{\parallel}}}\right|v_{the}),$
and 
\begin{equation}
Z(\zeta)=\frac{1}{\sqrt{\pi}}\int_{-\infty}^{+\infty}dt\frac{e^{-t^{2}}}{t-\zeta}
\end{equation}
is the plasma dispersion function \citep{plasma-disp}. 

We can replace Eq. (\ref{eq:Tkin}) in Eq. (\ref{eq:grnohmslawapp})
to obtain \citep{zocco:102309}
\begin{equation}
\left[\zeta^{2}-\frac{\tau}{Z}\frac{k_{\perp}^{2}d_{e}^{2}}{1-\Gamma_{0}\left(k_{\perp}^{2}\rho_{i}^{2}/2\right)}\right]\left[1+\zeta Z(\zeta)\right]=\frac{1}{2}k_{\perp}^{2}d_{e}^{2}.\label{eq:KAW}
\end{equation}
Looking for solutions with $\zeta=\omega/(\left|k_{\parallel}\right|v_{the})\ll1,$
one gets the dispersion relation of shear and kinetic Alfv\'en wave
\begin{equation}
\omega_{0}=\pm k_{\parallel}v_{A}k_{\perp}\rho_{i}\sqrt{\frac{1}{2}\left[\frac{Z}{\tau}+\frac{1}{1-\Gamma_{0}\left(k_{\perp}^{2}\rho_{i}^{2}/2\right)}\right]},
\end{equation}
where the damping rate is found by solving Eq. (\ref{eq:KAW}) perturbatively
in $\gamma/\omega_{0}\ll1,$ seeking for a solution $\omega=\omega_{0}+i\gamma:$
\begin{equation}
\gamma=-\left|k_{\parallel}\right|v_{A}\frac{k_{\perp}^{2}\rho_{i}^{2}}{4}\sqrt{\pi\frac{m_{e}}{m_{i}}\frac{Z^{3}}{\tau^{2}\beta_{e}}}.
\end{equation}

In the following, we solve analytically Eq. (\ref{eq:krehminhom})
by using Hermite polynomials as a basis in Hilbert space, and prove
analytically that the Hilbert space solution converges to Eq. (\ref{eq:Tkin})
when we take to infinity the number of Hermite moments kept in the
system.

\section{Hilbert Space}

We introduce the Hermite inverse transform defined as 
\begin{equation}
\hat{g}_{e}(v_{\parallel})=\sum_{m=2}^{\infty}\frac{H_{m}(\hat{v}_{\parallel})}{\sqrt{2^{m}m!}}\hat{g}_{m}F_{0e}(\hat{v}_{\parallel}^{2}),\label{eq:hilbert}
\end{equation}
 with coefficients 
\begin{equation}
\hat{g}_{m}=\frac{1}{n_{0e}}\int_{-\infty}^{\infty}d\hat{v}_{\parallel}\frac{H_{m}(\hat{v}_{\parallel})}{\sqrt{2^{m}m!}}\hat{g}_{e}(v_{\parallel}),
\end{equation}
 where $\hat{v}_{\parallel}=v_{\parallel}/v_{the},$ and $\hat{g}_{e}=2v_{the}^{-2}\int dv_{\perp}v_{\perp}\exp[-v_{\perp}^{2}/v_{the}^{2}]g_{e}.$
The resulting electron kinetic equation is 
\begin{equation}
\begin{split} & \frac{d}{dt}\hat{g}_{m}+v_{the}\hat{\mathbf{b}}\cdot\nabla\left(\sqrt{\frac{m+1}{2}}\hat{g}_{m+1}+\sqrt{\frac{m}{2}}\hat{g}_{m-1}-\delta_{m,1}\hat{g}_{2}\right)\\
 & =-\sqrt{2}\delta_{m,2}\hat{\mathbf{b}}\cdot\nabla u_{\parallel e}
\end{split}
\label{eq:exp}
\end{equation}
 Hence, for the first Hermite moments we obtain 
\begin{equation}
\begin{split} & \frac{d}{dt}\hat{g}_{2}+\sqrt{\frac{3}{2}}v_{the}\hat{\mathbf{b}}\cdot\nabla\hat{g}_{3}=-\sqrt{2}\hat{\mathbf{b}}\cdot\nabla u_{\parallel e},\end{split}
\label{eq:g2equation}
\end{equation}
 for $m=2,$ and 
\begin{equation}
\frac{d}{dt}\hat{g}_{m}+v_{the}\hat{\mathbf{b}}\cdot\nabla\left(\sqrt{\frac{m+1}{2}}\hat{g}_{m+1}+\sqrt{\frac{m}{2}}\hat{g}_{m-1}\right)=0,\label{eq:mmaggiroe4}
\end{equation}
 for $m\geq3.$ The moments $g_{0}$ and $g_{1}$ do not appear in
the summation in Eq. (\ref{eq:hilbert}) as they must be set to zero
in order to satisfy Eq. (\ref{eq:vincoli}).

Equation (\ref{eq:mmaggiroe4}) seems to suggest we should look for
an iterative solution. However, it manifests the typical problem of
a kinetic system, where low-order moments are coupled to high-order
ones, therefore requiring to solve an infinite number of equations
in order to know the full kinetic dynamics. We take advantage of the
scaling of Hermite coefficients with the Hermite order, and virtually
solve for an infinite number of equations in a very compact form.
We then prove that our solution is equivalent to the solution found
by using Landau contour integration.

\section{Landau-Hermite equivalence}

We find useful to consider the following limit

\begin{equation}
\frac{\hat{g}_{m}}{\hat{g}_{m-1}}\sim\frac{k_{\parallel}v_{the}}{\sqrt{m}\omega}\ll1,\label{eq:closMT-1}
\end{equation}
 with 
\begin{equation}
\frac{k_{\parallel}v_{the}}{\omega\sqrt{m}}\ll1,\,\,\,\mbox{and}\,\,\,\frac{k_{\parallel}v_{the}}{\omega}\sim1,
\end{equation}
which is true for large $m\gg1.$ 

If there is an $N\gg1$ for which $\hat{g}_{N+1}\ll\hat{g}_{N}$ in
the sense of Eq. (\ref{eq:closMT-1}), then the $Nth$ component
of the kinetic equation is 
\begin{equation}
\omega\hat{g}_{N}=k_{\parallel}v_{the}\sqrt{\frac{N}{2}}\hat{g}_{N-1}.\label{eq:closure}
\end{equation}
 We then use this solution for $\hat{g}_{N}$ in the equation for
the $N-1$ component and solve for the $N-1$ component as a function
of the $N-2$ component 
\begin{equation}
\left[\omega-k_{\parallel}v_{the}\frac{k_{\parallel}v_{the}N/2}{\omega}\right]\hat{g}_{N-1}=k_{\parallel}v_{the}\sqrt{\frac{N-1}{2}}\hat{g}_{N-2}.
\end{equation}
 After $n$ iterations we have 
\begin{equation}
\begin{split} & \hat{g}_{N-n}=k_{\parallel}v_{the}\sqrt{\frac{N-n}{2}}\hat{g}_{N-(n+1)}\frac{1}{\omega-k_{\parallel}v_{the}\frac{k_{\parallel}v_{the}(N-n+1)/2}{\omega-k_{\parallel}v_{the}\frac{k_{\parallel}v_{the}(N-n+2)/2}{\cdots-k_{\parallel}v_{the}\frac{k_{\parallel}v_{the}N/2}{\omega}}}}\end{split}
.\label{eq:gensolkin}
\end{equation}
 Equation (\ref{eq:gensolkin}) is a\textsl{ finite} continued fraction
that can be used to generate the \textsl{infinite} one which is the
exact solution of the collisionless problem.\textsl{ }Effects of finite
collisionality within this formalism have been considered somewhere
else \citep{sugama:2617,loureiro-scecco-zocco,PhysRevLett.111.175001,2014arXiv1407.1932P,2014arXiv1411.5604Z}.
Now, when $N-n=3,$ we are able to write $\hat{g}_{3}$ in Eq. (\ref{eq:g2equation})
explicitly as a function of all other $\hat{g}_{m}$ up to $\hat{g}_{N},$
and therefore the electron temperature perturbation in terms of the
continued fraction, which does not need to be truncated. The result
is 
\begin{equation}
\frac{v_{the}}{u_{\parallel e}}\frac{\delta T_{\parallel}}{T_{0e}}=\frac{k_{\parallel}}{\left|k_{\parallel}\right|}\frac{2}{\zeta}\frac{1}{1-\frac{1}{\zeta^{2}}\frac{3/2}{1-\frac{1}{\zeta^{2}}\frac{4/2}{1-\frac{1}{\zeta^{2}}\frac{5/2}{\ddots}}}}\equiv\frac{k_{\parallel}}{\left|k_{\parallel}\right|}\frac{2}{\zeta}\frac{1}{\mathcal{D}^{(2)}}.\label{eq:Tcontfrac}
\end{equation}
We now prove that Eq. (\ref{eq:Tcontfrac}) and (\ref{eq:Tkin}) are
the same.

We rewrite Eq. (\ref{eq:Tcontfrac}) in the following way
\begin{equation}
\frac{v_{the}}{u_{\parallel e}}\frac{\delta T_{\parallel}}{T_{0e}}=\frac{k_{\parallel}}{\left|k_{\parallel}\right|}\frac{1}{\zeta}\frac{1-\frac{1}{\zeta^{2}}\frac{2/2}{\mathcal{D}^{(2)}}-1}{-\frac{1}{2\zeta^{2}}},
\end{equation}
and notice that 
\begin{equation}
\frac{1}{\zeta}\frac{1-\frac{1}{\zeta^{2}}\frac{2/2}{\mathcal{D}^{(2)}}-1}{-\frac{1}{2\zeta^{2}}}=\frac{\frac{1}{\zeta}-\frac{1}{\zeta\mathcal{D}^{(1)}}}{-\frac{1}{2\zeta^{2}\mathcal{D}^{(1)}}},\label{eq:provefirst}
\end{equation}
where 
\begin{equation}
\mathcal{D}^{(1)}=\frac{1}{1-\frac{1}{\zeta^{2}}\frac{2/2}{1-\frac{1}{\zeta^{2}}\frac{3/2}{1-\frac{1}{\zeta^{2}}\frac{4/2}{\ddots}}}}.
\end{equation}
Similarly, we rewrite the RHS of Eq. (\ref{eq:provefirst}) as
\begin{equation}
\frac{\frac{1}{\zeta}-\frac{1}{\zeta\mathcal{D}^{(1)}}}{-\frac{1}{2\zeta^{2}\mathcal{D}^{(1)}}}=\frac{\frac{1}{\zeta}+2\zeta\left[1-\frac{1}{2\zeta^{2}\mathcal{D}^{(1)}}-1\right]}{1-\frac{1}{2\zeta^{2}\mathcal{D}^{(1)}}-1},
\end{equation}
which implies
\begin{equation}
\frac{v_{the}}{u_{\parallel e}}\frac{\delta T_{\parallel}}{T_{0e}}=\frac{k_{\parallel}}{\left|k_{\parallel}\right|}\frac{\frac{1}{\zeta\mathcal{D}^{(0)}}+2\zeta\left[1-\frac{1}{\mathcal{D}^{(0)}}\right]}{1-\frac{1}{\mathcal{D}^{(0)}}},
\end{equation}
with 
\begin{equation}
\mathcal{D}^{(0)}=\frac{1}{1-\frac{1}{\zeta^{2}}\frac{1/2}{1-\frac{1}{\zeta^{2}}\frac{2/2}{1-\frac{1}{\zeta^{2}}\frac{3/2}{\ddots}}}}.
\end{equation}
Since 
\begin{equation}
\frac{dZ(\zeta)}{d\zeta}=-2\left[1+\zeta Z\left(\zeta\right)\right],\label{eq:diffZ}
\end{equation}
with $Z(0)=i\sqrt{\pi},$ successive differentiation of Eq. (\ref{eq:diffZ})
yields \citep{McCabe}
\begin{equation}
\frac{Z^{(n)}}{Z^{(n-1)}}=\frac{-2n}{2\zeta+Z^{(n+1)}/Z^{(n)}},
\end{equation}
and therefore {[}see also Eq. 7.1.15 of \citet{abram}{]}
\begin{equation}
Z(\zeta)=-\frac{1}{\zeta\mathcal{D}^{(0)}}.
\end{equation}
Hence, we showed that
\begin{equation}
\frac{2}{\zeta}\frac{1}{\mathcal{D}^{(2)}}=-\frac{Z(\zeta)-2\zeta\left[1+\zeta Z(\zeta)\right]}{1+\zeta Z(\zeta)}.
\end{equation}
This concludes our proof.

\section{Conclusion}

We solved the problem of linear collisionless Landau damping for kinetic
and shear Alfv\'en waves by using both the traditional Laplace transform
approach and a Hermite transform.

We introduced a recursive formula for the coefficients of the inverse
Hermite transform that allowed us to construct a finite continued
fraction whose extension to infinite elements gave a new exact solution
for the electron distribution function. We proved that this new solution
is equivalent to the solution found by using Landau contour integration. 

\bibliographystyle{jpp}

\bibliography{Hilbert.bbl}
\end{document}